\documentclass[aps,twocolumn,epsf,floats,prl,superscriptaddress,nofootinbib]{revtex4-1}

\usepackage{graphics,graphicx,epsfig}
\usepackage{amssymb,color}
\usepackage{epsf,epstopdf,wrapfig}
\usepackage{amsmath}
\usepackage{amsmath}
\usepackage{amssymb}
\usepackage{graphicx}
\usepackage{bm}
\usepackage{xr}
\usepackage{scrextend}

\newcommand{\de}{\partial}

	\newcommand{\beq}{\begin{equation}}
	\newcommand{\eeq}{\end{equation}}
	\newcommand{\bea}{\begin{eqnarray}}
	\newcommand{\eea}{\end{eqnarray}}

\usepackage{lipsum}  
\newcommand{\gl}{\alpha}
\newcommand{\gj}{u}

\begin{document}

\title{Equilibrium to off-equilibrium crossover in homogeneous active matter}
\author{Andrea Cavagna}
\affiliation{Istituto Sistemi Complessi, Consiglio Nazionale delle Ricerche, UOS Sapienza, 00185 Rome, Italy}
\affiliation{Dipartimento di Fisica, Universit\`a\ Sapienza, 00185 Rome, Italy}

\author{Luca Di Carlo}
\affiliation{Dipartimento di Fisica, Universit\`a\ Sapienza, 00185 Rome, Italy}
\affiliation{Istituto Sistemi Complessi, Consiglio Nazionale delle Ricerche, UOS Sapienza, 00185 Rome, Italy}

\author{Irene  Giardina}
\affiliation{Dipartimento di Fisica, Universit\`a\ Sapienza, 00185 Rome, Italy}
\affiliation{Istituto Sistemi Complessi, Consiglio Nazionale delle Ricerche, UOS Sapienza, 00185 Rome, Italy}
\affiliation{INFN, Unit\`a di Roma 1, 00185 Rome, Italy}

\author{Tom\'as S. Grigera}
\altaffiliation[Permanent address: ]{Instituto de F\'\i{}sica de L\'\i{}quidos y Sistemas
  Biol\'ogicos (IFLySiB) --- CCT CONICET La Plata and Universidad
  Nacional de La Plata, Calle 59 no.\ 789, B1900BTE La Plata,
  Argentina}
\affiliation{Istituto Sistemi Complessi, Consiglio Nazionale delle Ricerche, UOS Sapienza, 00185 Rome, Italy}

\author{Giulia Pisegna}
\affiliation{Dipartimento di Fisica, Universit\`a\ Sapienza, 00185 Rome, Italy}
\affiliation{Istituto Sistemi Complessi, Consiglio Nazionale delle Ricerche, UOS Sapienza, 00185 Rome, Italy}

\begin{abstract}
We study the crossover between equilibrium and off-equilibrium dynamical universality classes in the Vicsek model near its ordering transition. Starting from the incompressible hydrodynamic theory of Chen et al \cite{chen2015critical}, we show that increasing the activity leads to a renormalization group (RG) crossover between the equilibrium ferromagnetic fixed point, with dynamical critical exponent $z = 2$, and the off-equilibrium active fixed point, with $z = 1.7$ (in $d=3$). We run simulations of the classic Vicsek model in the near-ordering regime and find that critical slowing down indeed changes with activity, displaying two exponents that are in remarkable agreement with the RG prediction. The equilibrium-to-off-equilibrium crossover is ruled by a characteristic length scale beyond which active dynamics takes over. Such length scale is smaller the larger the activity, suggesting the existence of a general trade-off between activity and system's size in determining the dynamical universality class of active matter. 
\end{abstract}

\maketitle

Despite several similarities with equilibrium statistical physics systems, active matter is known to display numerous off-equilibrium traits that make it qualitatively different from its non-active equilibrium siblings \cite{ginelli2016physics,toner_review}. In a certain class of systems, off-equilibrium effects are due to the interplay between an effective alignment interaction and a dynamical rewiring of the interaction network, which can give rise to a non-Hamiltonian coupling between the density and velocity fields \cite{toner_review, marchetti_review}. A classic example of such coupling is given by the Vicsek model \cite{vicsek+al_95,ginelli2016physics,chate+al_08} and its coarse-grained counterpart, the Toner and Tu field theory \cite{toner+al_95,toner+al_98,toner_12}, in which heterogeneous density structures emerge due to the density-velocity feedback \cite{chate+al_08b,marcehtti_fluctuations2010}. Density-velocity coupling is crucial in turning an equilibrium second-order transition into an off-equilibrium first-order one \cite{ginelli_onset2004, Bertin1, ginelli_metricfree2010}. It would therefore seem reasonable to assume that the coupling between density and velocity stays at the very core of off-equilibrium dynamics in active matter.

In fact, active motion and density-velocity coupling are related but distinct phenomena. The fact that particles continuously enter and exit the alignment interaction range of any given particle could violate detail balance also in a system with homogeneous density. Therefore, a question arises: Can activity lead to relevant off-equilibrium dynamics even in absence of any significant coupling between velocity and density, and therefore in absence of heterogeneous density structures?  If the answer to this question is positive, a second  problem consequently emerges, namely that of the equilibrium-to-off-equilibrium crossover. In a system  with strong feedback between velocity and density, switching off activity may freeze the interaction network into a variety of complex patterns (bands, droplets, accumulation points, etc); dynamics may strongly depend on such specific configurations, thus voiding the very notion of universality class. On the other hand, in an active yet homogeneous system the limit of zero activity gives rise to an equally homogeneous interaction network, making the dynamics uniquely converge to an equilibrium universality class. In this second case, assessing how activity changes the equilibrium universality class is a well-defined problem. 

Here we approach these two issues in the Vicsek model \cite{vicsek+al_95}, focusing on its scaling region just above the ordering transition. 
The near-ordering phase of the Vicsek model has been shown to be a very sound theoretical candidate to describe natural swarms of 
insects \cite{attanasi2014collective, attanasi2014finite}, for which experiments are starting to provide quantitative characterizations of dynamical universality classes \cite{cavagna2017dynamic}, thus making a comparison between theory and experiments compelling. 
The near-ordering phase is also the ideal arena where to investigates the questions above: the near-ordering Vicsek regime is less prone to developing heterogeneities than the ordered phase \cite{baglietto2009nature, baglietto2008finite}, suggesting that the coupling between velocity and density, and therefore perhaps activity, is not relevant there; yet, experiments on real swarms give dynamical critical exponents far from their equilibrium counterpart \cite{cavagna2017dynamic}, suggesting that activity {\it is} in fact relevant in this phase, despite its homogeneity.  

Our strategy to investigate this contradiction will be the following: first, we will perform a renormalization group study of the hydrodynamic theory of active ordering under the assumption of {\it incompressibility}, in such a way as to inhibit the coupling between density and velocity fluctuations from the outset, and check under what conditions activity is relevant in this homogenous case. Secondly, we will perform numerical simulations of the Vicsek model in the near-critical phase and verify that the theoretical results about the incompressible case apply also {\it without} imposing incompressibility in the simulation. We will find that activity-driven off-equilibrium effects are significant also in absence of velocity-density coupling, giving rise to an equilibrium to off-equilibrium crossover between two different dynamical universality classes.

\begin{figure}
\includegraphics[width=0.5 \textwidth,trim={0.5cm 0 0 0},clip]{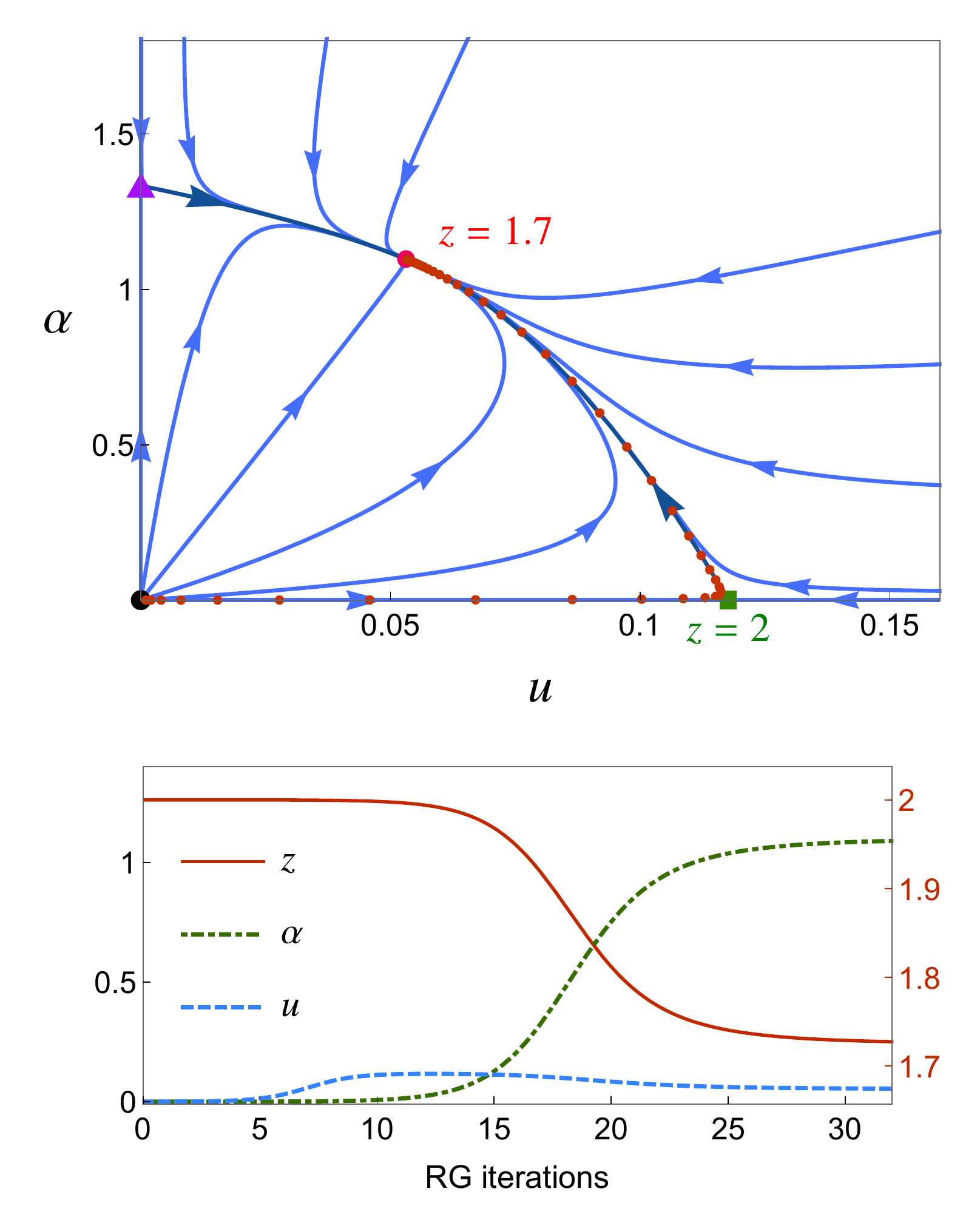}
\caption{{\bf RG flow and crossover.} (Top) The system admits four fixed points: The free theory, $\gl^* = 0$, $\gj^* = 0$ (Black disk) and the incompressible stirred fluid $\gl^* = 4/3 \epsilon$, $\gj^*  = 0$ (purple triangle \cite{forster1977large}), are of no interest to us. On the other hand, the equilibrium ferromagnet, $\gl^* = 0$, $\gj^* = 2/17 \epsilon$ (green square), and the Vicsek active-ferromagnetic fixed point, $\gl^* = 124/113 \epsilon $, $\gj^* = 6/113 \epsilon $ (red disk), give rise to the equilibrium to off-equilibrium crossover. If the initial value of the activity, $\gl_0$, is small, the flow quickly approaches the equilibrium fixed point (red dots) and remains around it for many RG iterations, before crossing over to the Vicsek fixed point. (Bottom) The crossover is evident when plotting the running parameters over an RG streamline with small $\gl_0$. The running critical exponent $z$ flows from its equilibrium value $z=2$ to its off-equilibrium value, $z= 1.7$ (in $d=3$). 
}
\label{fig:FluxDiagram}
\end{figure}

Inspired by the classic calculation of \cite{forster1977large} on stirred turbulence, the incompressible hydrodynamic theory of the Vicsek model near its ordering transition has been developed in \cite{chen2015critical}, whose results we briefly summarize here. Incompressibility implies constant density, hence one can work solely with the velocity field, $\boldsymbol{v}(\boldsymbol{x},t)$, which is governed by the equation \cite{chen2015critical},
\begin{equation}
    \frac{\partial {\boldsymbol v}}{\de t}  + \lambda_0 (\boldsymbol v \cdot \nabla) \boldsymbol v = \Gamma_0\nabla^2 \boldsymbol{v} + J_0 v^2 \boldsymbol{v}+ \boldsymbol{f} + \boldsymbol {\nabla} P \ , 
    \label{eq:EOM}
\end{equation}
where $\boldsymbol{f}$ is a white noise with variance $2\tilde \Gamma_0$; at equilibrium the kinetic coefficient $\Gamma_0$ would be equal to $\tilde \Gamma_0$, but the self-propulsion term, $(\boldsymbol v \cdot \nabla) \boldsymbol v$, can violate this condition.  The pressure $P$ enforces incompressibility, $\nabla \cdot \boldsymbol{v} = 0$ \cite{forster1977large}. The coupling $J_0$ governs the Landau-Ginzburg  ferromagnetic interaction \cite{cardy1996scaling}, while the linear mass term has been set to zero as we work at criticality. The Renormalization Group (RG) analysis of \cite{chen2015critical} showed that the theory has two effective coupling constants that become irrelevant above the upper critical dimension, $d_c=4$; the first one encodes the strength of activity, $\gl_0 =\lambda_0^2 \, (\tilde \Gamma_0/\Gamma_0^{3} ) \Lambda^{-\epsilon}$, while the second is proportional to the ferromagnetic coupling, $\gj_0 =  J_0 \, (\tilde \Gamma_0/ \Gamma_0^{2}) \Lambda^{-\epsilon}$, where $\epsilon = 4-d$ and $\Lambda$ is the cutoff. The RG equations at order $\epsilon$ (one loop) are \cite{chen2015critical},
\begin{equation} 
\begin{split}
    \gl_{l+1} &=\gl_l \; b^\epsilon ( 1- 3/4 \  \gl_l \ln b - 10/3 \  \gj_l \ln b ) 
     \\ 
    \gj_{l+1} &= \gj_l  \; b^\epsilon   ( 1- 1/2 \ \gl_l\ln b-17/2 \ \gj_l\ln b  )   
    \end{split}
    \label{flow}
\end{equation}
where $b>1$ is the RG rescaling factor \cite{wilson1971renormalization1}. 
The RG flow and fixed points are displayed in Fig.\ref{fig:FluxDiagram}. The fixed point with $\alpha^*=0$ and $u^*\neq0$ (green square), describes an equilibrium ferromagnet, and it has the classic dynamical critical exponent $z = 2$ (at one loop) \cite{hohenberg1977theory};
this equilibrium fixed point is stable along the axis $\gl=0$, but as soon as activity is switched on, the unstable direction leads the flow to the only stable fixed point, where both ferromagnetism {\it and} activity are relevant, $\alpha^*\neq 0$,  $u^*\neq0$ (red dot). At this off-equilibrium fixed point the dynamical critical exponent  changes significantly, $z = 1.7$ in $d=3$, defining a new dynamical universality class \cite{chen2015critical}.

\begin{figure*}
\centering
\includegraphics[width=1.0 \textwidth]{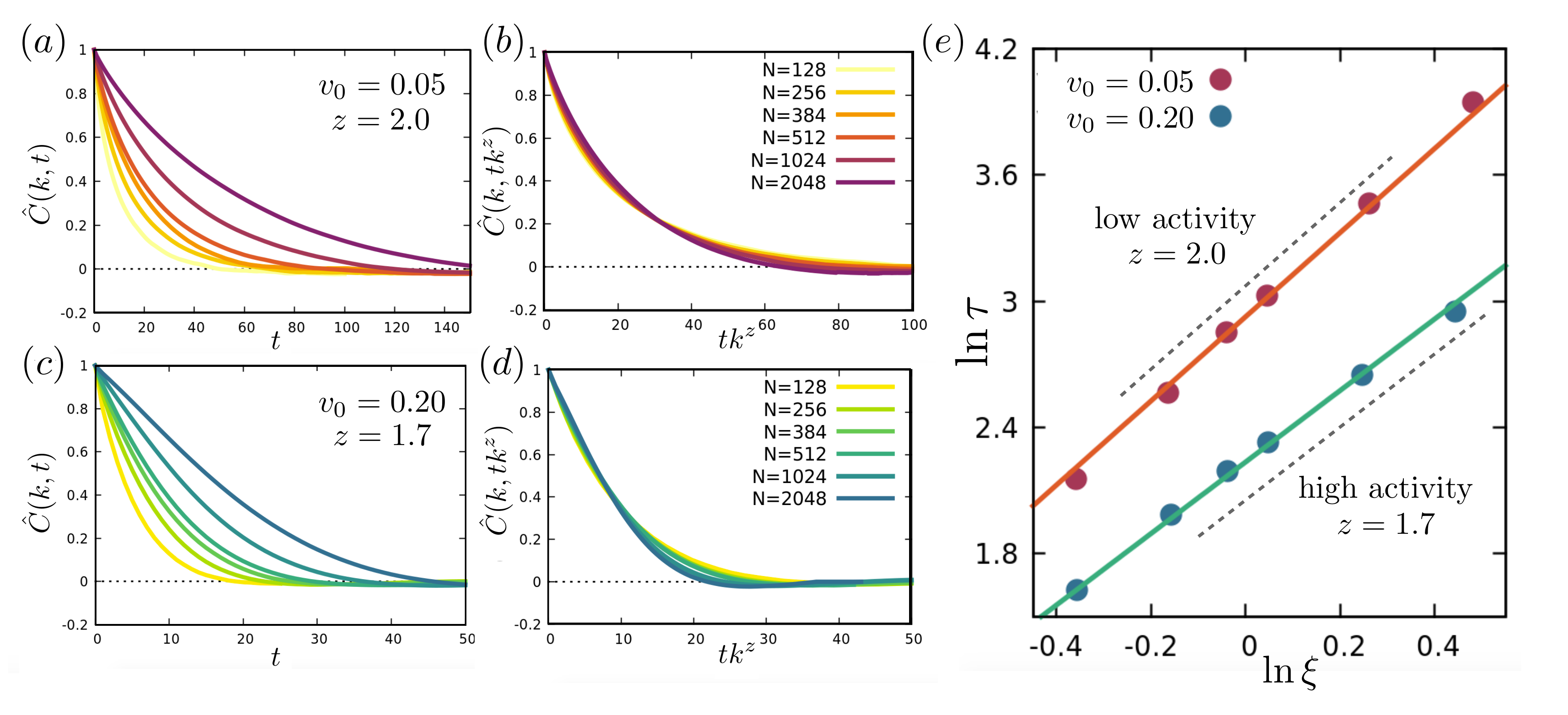}
\caption{{\bf Vicsek dynamics in three dimensions.} $(a)$, $(c)$ Normalized dynamical correlation functions, $\hat C(k,t) = C(k,t)/C(k,0)$, at $k=1/\xi$ for $v_0=0.05$ and $v_0=0.20$, respectively. $(b)$, $(d)$ Correlation functions plotted against the scaling variable $tk^z$; $z \simeq 2$ in $(b)$ and $z\simeq 1.7$ in $(d)$. $(e)$ Relaxation time vs correlation length for the two different values of the speed $v_0$. Lines are best fit to the RG results,  $z=1.7$ (blue, high $v_0$, high activity) and $z=2$ (red, low $v_0$, low activity).}
\label{simu}
\end{figure*}

The idea that we develop in this work is that the interplay between the equilibrium and the off-equilibrium RG fixed points  contains the key to understanding to what extent activity is relevant in realistic systems, with finite size and finite activity. An analysis of equations \eqref{flow} shows that, if the initial RG value of the activity coupling constant, that is the {\it physical} value $\gl_0$, is small, the flow quickly reaches the equilibrium fixed point and it remains in the vicinity of it {\it for a large number of RG iterations}, before eventually crossing over to the active, off-equilibrium fixed point (see Fig.\ref{fig:FluxDiagram}). Accordingly, the running value of the dynamical critical exponent along the flow ($z_l = 2-1/4 \ \gl_l$ - see Supp. Info.), abruptly changes from its equilibrium value ($z=2$) to its off-equilibrium value ($z=1.7$, in $d=3$). To understand how this RG crossover impacts on the critical dynamics of the system, we must turn our attention to the correlation length.

Close to criticality, the physical correlation length, $\xi$, is large, but it gets reduced by a factor $b$ at each RG iteration, $\xi_{l+1}=\xi_l/b=\xi/b^l$. Once the running correlation length, $\xi_{l+1}$, becomes of order $1/\Lambda$, the system is no longer near-critical and the flow must stop; this condition implies $b^{l_\text{STOP}} = \Lambda \xi$ \cite{cardy1996scaling}. If when the RG flow stops we are still in the neighbourhood of the {\it equilibrium} fixed point, the exponent $z=2$ will rule the critical dynamics of the system, even in presence of active terms in the equations of motion. By expanding equations \eqref{flow} in the vicinity of the equilibrium fixed point, we find that the activity coupling evolves as $\gl_{l+1} = \gl_l b^{\kappa} = \gl_0 b^{\kappa l}$, with scaling dimension $\kappa = (31/51) \epsilon$. Because the {\it off-equilibrium} fixed point has $\gl^*=\mathrm O(\epsilon)$, asking that activity is still irrelevant when we have left the critical region amounts to asking $\gl_{\mathrm{STOP}} \ll \epsilon$, which gives, $\gl_0 b^{\kappa l_\mathrm{STOP}}=\gl_0 (\Lambda\xi)^{\kappa} \ll \epsilon$. If we now define the crossover length scale,
\beq
{\cal R}_c= \left(\epsilon/\gl_0 \right)^\frac{1}{\kappa}\, \Lambda^{-1} \ ,
\eeq
we conclude that that for $\xi \ll {\cal R}_c$ equilibrium dominates, and $z=2$; conversely, for $\xi \gg {\cal R}_c$ dynamics is ruled by the active off-equilibrium fixed point, and $z=1.7$ (in $d=3$). The crossover length scale decreases for increasing activity, hence shrinking the equilibrium regime; eventually, if $\gl_0$ is so large that ${\cal R}_c < 1/\Lambda$ (the microscopic scale) one observes off-equilibrium critical dynamics at all macroscopic scales. Conversely, when activity decreases, the range over which the equilibrium dynamics dominates broadens, and this has an interesting implication in finite-size systems: if size and activity are such that ${\cal R}_c > L$, we cannot reach a scale where we observe the effects of activity, so that equilibrium dominates critical dynamics at {\it all} physical scales. This finite-size crossover as a function of activity is what we test in numerical simulations.

In the Vicsek model \cite{vicsek+al_95} all particles have the same speed, $v_0$, hence it is convenient to define the velocities as ${\bm v}_i = v_0\,\bm{\varphi}_i$, and write the equations in terms of the unitary orientations, $\bm{\varphi}_i$,
\begin{equation}
\begin{split}
\bm{\varphi}_i(t+1) &= \mathcal R_\eta \left(   \sum_j n_{ij}(t) \bm{\varphi}_j (t) \right) 
\\
\bm{r}_i (t+1) &= \bm{r}_i(t) +v_0\, \bm{\varphi}_i (t+1) \ ,
\end{split}
\label{vicsek}
\end{equation}
where $n_{ij}(t)$ indicates the metric interaction network \cite{vicsek+al_95,ginelli2016physics}, which changes in time due to the active motion of the particles. The noise operator $\mathcal R _{\eta}$ normalizes and rotates its argument randomly by an angle $4 \pi \eta$. We simulate systems in $d=3$ with $N=128, 256, 384, 512, 1024, 2048$ particles with PBC, and metric interaction range $r_c = 1$. Simulations are run at constant noise, $\eta=0.45$, while the density is tuned in order to follow the ordering transition at various $N$ and therefore explore the near-critical scaling regime (see Supp. Info. for details). We stress that, although we will compare simulations with RG results on the incompressible theory, we {\it do not} impose incompressibility in the model; we merely monitor homogeneity and verify the absence of significant density fluctuations in the phase that we consider (see Supp. Info).

In the microscopic model \eqref{vicsek} the equilibrium limit, $v_0\to 0$, is transparent: the second equation reduces to $\bm{r}_i (t+1) = \bm{r}_i(t)$, so that the interaction network is frozen, $n_{ij}(t)=n_{ij}$, and the first equation becomes that of an equilibrium ferromagnet. This suggests that we can tune activity by simply tuning the speed. However, the coarse-grained equation \eqref{eq:EOM} and the effective coupling constant, $\gl_0 =\lambda_0^2 \, (\tilde \Gamma_0/\Gamma_0^{3} ) \Lambda^{-\epsilon}$, actually governing activity in the theory, have an unclear limit for $v_0\to 0$. To make progress, it is convenient to separate also the coarse-grained velocity as, $\boldsymbol v(x) = v_0 \boldsymbol\varphi(x)$, so that the microscopic speed $v_0$ appears as an explicit parameter of the field theory (note that $\boldsymbol\varphi(x)$ is {\it not} a unitary field).  By comparing the microscopic with the coarse-grained parameters, one can prove (see Supp. Info.) that the strength of the noise, $\tilde\Gamma_0$, is of order $v_0^2$ in the limit $v_0\to 0$, while the other parameters are non-singular; hence, the activity coupling constant scales as, $\gl_0 \sim v_0^2$. Changing speed $v_0$ in the microscopic model is therefore equivalent to changing the activity coupling constant in the hydrodynamic theory. In order to test the crossover we run simulations at $v_0=0.05$ (low activity) and $v_0= 0.2$ (high activity).

To determine the dynamical critical exponent, $z$, we calculate the dynamical correlation function \cite{cavagna2018physics},
\begin{equation}
C(k,t) = \biggl<  \frac{1}{N} \sum_{ij} \delta \bm v_i (t_0) \cdot \delta \bm v_j (t_0+t) \, \frac{\sin (k r_{ij})}{k r_{ij}}  \biggr>_{t_0} \ ,
\end{equation}
where $r_{ij}=|\bm r_i(t_0)-\bm r_j(t)|$. The normalized correlation obeys dynamical scaling, $\hat C(k,t)= C(k,t)/C(k,0) = f(t/\tau(k); k\xi)$, where the $k$-dependent relaxation time is an homogeneous function, $\tau(k)= k^z g(k\xi)$ ($f$ and $g$ are scaling functions) \cite{HH1969scaling}. 
The first relation implies that when we select $k=1/\xi$, we have $\hat C(k,t)= F(k^zt)$, namely different correlation functions collapse onto each other once plotted against the scaling variable $k^zt$.
The second relation can be rewritten as,  $\tau(k)= \xi^z \tilde g(k\xi)$, so that the $k=0$ (i.e. collective) relaxation time grows with the correlation length as $\tau = \xi^z$, the phenomenon known as critical slowing down \cite{HH1969scaling}. We employ both relations to estimate $z$; results are reported in Fig.~\ref{simu}.

In the low activity regime, $v_0=0.05$, we find that critical slowing down is ruled by the equilibrium RG exponent, $z = 2.0$; consistently, the respective dynamical correlation functions collapse when time is rescaled as $tk^2$ (Fig.~\ref{simu}, panels a, b, and e). This result is consistent with simulations performed at the same value of $v_0$ in \cite{cavagna2017dynamic}. On the other hand, when we change the speed to $v_0=0.2$, the system indeed crosses-over to the active dynamical universality class, characterized by the off-equilibrium RG critical exponent $z=1.7$ (Fig.~\ref{simu}, panels c, d and e). Given the limited span in the correlation length and relaxation time, we have run three different statistical tests to check the compatibility between data and different fits; all three tests support the validity of the RG exponents in reproducing the numerical data (see Supp. Info.). As it often happens with RG, such a good agreement between experiments and theory is quite astonishing, considering that results in $d=3$ are obtained by setting $\epsilon=1$ in an expansion for small $\epsilon$. Fortunately, people more authoritative than us have thoroughly discussed such a miracle \cite{wilson1974renormalization}. Yet, if we further reflect on the fact that here the RG is perfectly capturing a {\it crossover} between two different universality classes, one of which is {\it out of equilibrium}, we cannot help being more than a bit nonplussed.

In conclusion, the crossover between equilibrium and off-equilibrium universality classes that emerges in the hydrodynamic theory of active matter of \cite{chen2015critical}, is not an artifact of incompressibility, as it is fully supported by numerical simulations of the standard Vicsek model, in which incompressibility is {\it not} enforced. This demonstrates that activity is able to modify the critical exponents even in absence of a feedback between density and velocity. At the moment, we cannot tell whether including this feedback would bring the system to yet another non-equilibrium universality class or would merely impair the rule of RG, thus leading the system far from any notion of universality.  More work is needed to settle that question. In relation to the study of real biological systems, though, the present results are quite encouraging. Bird flocks and insect swarms display very weak density fluctuations \cite{mora2016local}, so that the incompressible hydrodynamic theory and the equilibrium to non-equilibrium crossover that it encapsulates, seem the appropriate starting point to more complex endeavours, as the inclusion of non-dissipative mode-coupling terms between velocity and spin, which seem necessary to match theory with experiments \cite{cavagna2019dynamical,cavagna2019renormalizationl,cavagna2017dynamic}.

We thank M. Cencini, H. Chat\'e, F. Ginelli, and S. Melillo for discussions. This work was supported by ERC Grant RG.BIO (785932) to AC, and ERANET-CRIB grant to AC and TSG.  TSG was partially supported by CONICET and Universidad Nacional de La Plata (Argentina).

\bibliographystyle{apsrev4-1}
\bibliography{general_cobbs_bibliography_file}

\vfill
\eject
\newpage

. 

\vfill
\eject
\newpage

\section{Supplementary Information}

\subsection{Fixed points coupling constants and dynamical critical exponent}

The RG fixed points rule the critical dynamics of the system, indeed critical exponents can be derived using recursion relations of the parameters and the fixed point values of the effective coupling constants. At order $\epsilon$ (one loop) the recursion relations are: 
\begin{align}
 \Gamma_{l+1} &= \Gamma_l \ b^{z_l -2}( 1 -\frac 14  \alpha_l \ln b)\\ 
 \tilde \Gamma_{l+1} &= \tilde \Gamma_l \ b^{z_l-d-2 \chi_l} \\
   \gl_{l+1} &=\gl_l \; b^\epsilon ( 1- 3/4 \  \gl_l \ln b - 10/3 \  \gj_l \ln b ) 
     \\ 
    \gj_{l+1} &= \gj_l  \; b^\epsilon   ( 1- 1/2 \ \gl_l\ln b-17/2 \ \gj_l\ln b  )   
\end{align}  

The running  dynamical critical exponent $z_l$ and the scaling dimension of the field  $\chi_l$ are fixed imposing that both the natural time scale of the model  $\Gamma_l$ and the noise strength $\tilde \Gamma_l$ remain untouched by the RG flow; namely imposing $\Gamma_{l+1} = \Gamma_l$ and $\tilde \Gamma_{l+1} = \tilde \Gamma_l$.
\begin{equation}
z_l = 2 - \frac{1}{4} \alpha_l  \qquad \chi_l = \frac{ z_l-d}{2}
\end{equation}

 It is possible to show that this quantity $z_l$ entirely rules the critical dynamics of the system. The temporal correlation function $C(t,\xi,k,\mathcal P)$ satisfies the following RG recursion relation, \begin{equation}
 C(t,\xi,k, \mathcal P_{0}) = b^{2 l z_l} C(k b,\xi/b, t b^{-l z_l }, \mathcal P_l)
 \label{scaling}
\end{equation}
where $\mathcal P_l$ represents the set of parameters of the model flowing towards the fixed point. Using the RG flow stop condition $\Lambda \xi_{l_\text{STOP}} = 1$, which implies $b^{l_\mathrm{STOP}} = \Lambda \xi$, the scaling relation \eqref{scaling} becomes
\begin{equation}
	C(t,\xi,k,\mathcal P) =(\Lambda \xi)^{2 z_{l_\mathrm{STOP}}}C( t/\xi^{z_{l_\mathrm{STOP}}},\Lambda^{-1}, \Lambda k \xi, \mathcal{P}^*) 
\end{equation}
At fixed $k \xi$ the correlation function depends only on the ratio $t/\xi^{z_{l_\mathrm{STOP}}}$, hence the relaxation time $\tau$, ruling the decay of the temporal correlation function, must obey
\begin{equation}
 \tau \sim \xi^{z_{l_\mathrm{STOP}}}.
\end{equation}

The value of $z_{l_\mathrm{STOP}}$ depends on the fixed point value of the effective coupling $\alpha_l$ reached when the RG flow stops, or more in general on which of the two fixed points is approached at the end of the RG flow. If the RG flow stops in the neighborhood of the $z = 2$ \textit{equilibrium} fixed point, then $\tau \sim \xi^2$; conversely if the RG flow stops in the neighborhood of the \textit{non-equilibrium} $z= 1.7$ fixed point we have $\tau \sim \xi^{1.7}$.

\subsection{Microscopic speed tunes hydrodynamic activity}

To reproduce in the microscopic model the crossover from equilibrium/inactive to off-equilibrium/active dynamics that we have discovered in the coarse-grained theory, we have to tune some microscopic parameter whose effect is to change the activity coupling constant, $\gl =\lambda^2 \, (\tilde \Gamma/\Gamma^{3} ) \Lambda^{-\epsilon}$ (to lighten the notation we drop here all the 0 subscripts from the bare parameters). The intuitive candidate to tune is the speed $v_0$, since for $v_0=0$ the Vicsek model reduces to an equilibrium ferromagnet. However, it is unclear what is the effect of the limit $v_0\to 0$ in the hydrodynamic theory.  We develop here a method to find the dependence of the coarse-grained parameters on microscopic speed in the limit of small $v_0$. 

The original Vicsek model \cite{vicsek+al_95} is formulated in terms of the microscopic velocities, 
\begin{equation}
\begin{split}
\bm{v}_i(t+1) &= \mathcal R_\eta \left(   \sum_j n_{ij}(t) \bm{v}_j (t) \right) 
\\
\bm{r}_i (t+1) &= \bm{r}_i(t) +  \bm{v}_i (t+1) \ ,
\end{split}
\label{vicovaro}
\end{equation}
with the constraint of fixed speed, $|\bm{v}_i| = v_0$. The corresponding incompressible hydrodynamic theory is \cite{chen2015critical}, 
\begin{equation}
\begin{split}
    \frac{\partial {\boldsymbol v}}{\de t}  + \lambda (\boldsymbol v \cdot \nabla) \boldsymbol v =& \Gamma \nabla^2 \boldsymbol{v} + 
    (a+ J v^2) \boldsymbol{v}+ \boldsymbol{f} 
    \\
    \langle \boldsymbol{f}(x,t)\boldsymbol{f}(x',t')\rangle =& 2 \tilde\Gamma \ \delta^{(d)}(x-x')\delta(t-t')
    \end{split}
    \label{burgio}
\end{equation}
where we have reinstated the mass term, $a \boldsymbol{v}$, for the sake of generality. The coarse-grained parameters of the hydrodynamic theory $(\lambda, \Gamma, J, a, \tilde\Gamma)$ depend in complicated ways on the microscopic parameters $v_0, \eta, r_c$, and also on the specific coarse-graining procedure.  Working out the precise connection between the two sets of parameters is highly nontrivial (see however \cite{bertin2006boltzmann, peshkov2014boltzmann}). Even the simpler task of determining the dependence of the coarse-grained parameters on the microscopic speed, $v_0$, is difficult, because the limit $v_0\to 0$ is hard to handle in \eqref{vicovaro} and \eqref{burgio}. To get round this difficulty, we rewrite the microscopic equations of the Vicsek model in terms of the unitary orientations, as we did in the main text, 
\begin{equation}
\begin{split}
\bm{\varphi}_i(t+1) =& \mathcal R_\eta \left(   \sum_j n_{ij}(t) \bm{\varphi}_j (t) \right) 
\\
\bm{r}_i (t+1) =& \bm{r}_i(t) +v_0\, \bm{\varphi}_i (t+1) \ ,
\end{split}
\label{ramoracce}
\end{equation}
To write a hydrodynamic equation for this model, we write the coarse-grained velocity field as, $\bm{v}(x,t) = v_0 \bm{\varphi}(x,t)$, where we have introduced the coarse-grained ``orientation" field, $\bm{\varphi}(x,t)$. It is important to note that, while $\bm{\varphi}_i$ are unitary vectors, the field $\bm{\varphi}(x,t)$ is {\it not}, as an effect of the coarse-graining; this is why we write ``orientation" in quotes. The coarse-grained velocity can be zero for two reasons: {\it i)} because the system is misaligned, so that the local average gives a zero value; {\it ii)} because the microscopic speed $v_0$ is zero; by decomposing the velocity as $\bm{v}(x,t) = v_0 \bm{\varphi}(x,t)$ we decouple these two effects, so that $\bm{\varphi}(x,t)$ can be zero only of the first reason. By using the same exact arguments that lead from \eqref{vicovaro} to \eqref{burgio}, we can say that \eqref{ramoracce} leads to,
\begin{equation}
\begin{split}
    \frac{\partial \bm{\varphi}}{\de t}  + \lambda' (v_0 \bm{\varphi} \cdot \nabla) \bm{\varphi} =&  \Gamma' \nabla^2\bm{\varphi} + 
    (a'+J' \varphi^2) \bm{\varphi} + \boldsymbol{f'} 
    \\
    \langle \boldsymbol{f'}(x,t) \boldsymbol{f'}(x',t') \rangle =& 2 {\tilde\Gamma'} \ \delta^{(d)}(x-x')\delta(t-t')
    \end{split}
    \label{donnasummer}
\end{equation}
In this last theory, apart from the usual implicit dependence of all parameters on the microscopic ones, there is also an {\it explicit} dependence on $v_0$, coming from the fact that the material derivative must contain a term $(\boldsymbol v \cdot \nabla)$ irrespective of the field that gets transported, be it $\boldsymbol v$ or $\bm{\varphi}$. We can now connect the parameters of the two hydrodynamic theories, that for the velocity and that for the orientation. To do so, let us multiply by $v_0$ both sides of equation \eqref{donnasummer}, and recall that $v_0 \bm{\varphi} = \bm{v}$, thus yielding,
\beq
\frac{\partial \bm{v}}{\de t}  
+ \lambda' (\bm{v} \cdot \nabla) \bm{v} =  \Gamma' \nabla^2\bm{v} + 
\left(a'+\frac{J'}{v_0^2} v^2\right)\bm{v} + v_0\boldsymbol{f'} 
\label{sandramilo}
\eeq
This equation must be equal to \eqref{burgio}, because coarse-graining the microscopic orientations $\bm{\varphi}_i$ and then multiplying by $v_0$, must give the same field theory as coarse-graining directly the microscopic velocities, $\bm{v}_i$, as long as $v_0$ does not fluctuate. Hence, we can read the parameters of the hydrodynamic theory for $\bm v$ in terms of those of $\bm\varphi$, 
\begin{equation}
\begin{split}
\lambda =& \lambda' 
\\
\Gamma=&\Gamma'
\\
a =& a'
\\
J =& J'/v_0^2 
\\
\tilde\Gamma =& v_0^2 {\tilde\Gamma'}
\end{split}
\label{barbapapa}
\end{equation}
The great advantage at this point is that the primed parameters have a simple, well-defined limit for $v_0\to0$, because in this limit the microscopic model \eqref{ramoracce} becomes an equilibrium ferromagnet\footnote{In drawing this conclusion we are assuming that the interaction network gets frozen by the $v_0=0$ condition into a reasonably homogeneous configuration, where standard equilibrium statistical physics applies, which is all the more plausible when we work in the incompressible case, as in the present work.} and therefore in this same limit the corresponding hydrodynamic theory, eq. \eqref{donnasummer}, becomes that of an equilibrium ferromagnetic field theory, normally called Model A, which can schematically be written as \cite{hohenberg1977theory,cardy1996scaling},
\begin{equation}
    \frac{\partial \bm{\varphi}}{\de t}  = \Gamma_\mathrm{eq}\frac{\partial {\cal H}}{\partial \bm{\varphi}}
    + \boldsymbol{\zeta} 
\quad \quad , \quad\quad
    \langle \boldsymbol{\zeta} \boldsymbol{\zeta} \rangle = 2 \Gamma_\mathrm{eq}
    \label{suka}
\end{equation}
where $\Gamma_\mathrm{eq}$ is the equilibrium kinetic coefficient, and $\cal H$ is the classic Landau-Ginzburg Hamiltonian, 
\beq
{\cal H} = \int d^dx\  (\nabla \bm{\varphi})^2 + r_\mathrm{eq}\varphi^2 + u_\mathrm{eq} \varphi^4
\label{LG}
\eeq
where $r_\mathrm{eq}$ is the equilibrium (square) mass and $u_\mathrm{eq}$ is the equilibrium non-gaussian coupling constant. By comparing \eqref{donnasummer}, with \eqref{suka} and \eqref{LG}, we can determine the $v_0\to 0$ limit of the parameters of the coarse-grained theory for the ``orientation" field $\bm{\varphi}$,
\begin{equation}
\begin{split}
a'(v_0=0) =& \Gamma_\mathrm{eq} \; r_\mathrm{eq}
\\
J'(v_0=0) =& \Gamma_\mathrm{eq} \; u_\mathrm{eq} 
\\
\Gamma'(v_0=0) =& \Gamma_\mathrm{eq}
\\
\tilde\Gamma'(v_0=0) =& \Gamma_\mathrm{eq}
\end{split}
\label{assolto!}
\end{equation}
This means that the coarse-grained parameters of the hydrodynamic equations for the ``orientation'' field reach a non-singular well-defined value in the equilibrium limit $v_0\to 0$. The only exception is the material derivative coupling constant, $\lambda'$, that remains undetermined, as it should, due to the fact that the explicit $v_0$ factor in front of it drives it to zero in the $v_0\to 0$ limit. We can now use \eqref{assolto!} and \eqref{barbapapa} to work out the small $v_0$ behaviour of the original coarse-grained coupling constants in terms of the speed and the equilibrium parameters,
\begin{equation}
\begin{split}
a &\sim \Gamma_\mathrm{eq} \; r_\mathrm{eq}
\\
J &\sim \Gamma_\mathrm{eq} \; u_\mathrm{eq}/v_0^2 
\\
\Gamma&\sim\Gamma_\mathrm{eq}
\\
\tilde\Gamma &\sim v_0^2 \;  \Gamma_\mathrm{eq} 
\end{split}
\label{ugotognazzi}
\end{equation}
The last equation is the most important to us: the noise variance in the hydrodynamic theory of the velocity field, $\tilde \Gamma$, must go to zero as $v_0^2$ when the speed goes to zero, in order to provide the correct equilibrium limit of the microscopic model. The RG activity coupling constant is $\gl =\lambda^2 \, (\tilde\Gamma /\Gamma^{3} ) \Lambda^{-\epsilon}$, and therefore its small $v_0$ behaviour is given by,
\beq
\gl =\frac{\lambda^2\,v_0^2}{\Gamma_\mathrm{eq}^2}  \Lambda^{-\epsilon} \sim v_0^2 \ ,
\label{apritisesamo}
\eeq
where we have simply assumed that the material derivative coupling constant, $\lambda$, does not diverge in the $v_0\to 0$ limit, as there is no physical reason for that to happen. 

We conclude that, in order to tune activity in the coarse-grained field theory, we can indeed tune the speed of the microscopic theory; the nonlinear exponent $2$ in \eqref{apritisesamo} also shows that activity grows rather rapidly with $v_0$, hence small changes of the speed can have great  effects on critical dynamics, as our simulations confirms. Notice that this apparently intuitive link between activity and speed goes through a not entirely trivial mechanism, namely the dependence on speed of the noise intensity in the original hydrodynamic theory for the velocity.

One final consideration is in order. From \eqref{ugotognazzi} we learn that the ferromagnetic coupling constant, $J$, of the hydrodynamic theory for the velocity, {\it diverges} for $v_0\to0$, which may seem troubling, if not outright wrong. In fact, we should remember that the {\it effective} ferromagnetic coupling constant, namely the parameter truly ruling the RG flow and fixed points, is $\gj =  J \, (\tilde \Gamma/ \Gamma^{2}) \Lambda^{-\epsilon}$. From \eqref{ugotognazzi}, we see that the $v_0\to0$ limit of this effective coupling constant is,
\beq
\gj \sim \frac{\Gamma_\mathrm{eq} u_\mathrm{eq}\; v_0^2\,\Gamma_\mathrm{eq}}{v_0^2\,\Gamma_\mathrm{eq}^2}\; \Lambda^{-\epsilon} = u_\mathrm{eq} \Lambda^{-\epsilon} 
\eeq
which is exactly the (finite) non-Gaussian effective coupling constant of the equilibrium Landau-Ginzburg theory, as it should. Hence, all is in order in the $v_0\to 0$ limit. Physically, what happens is that the ferromagnetic term in the hydrodynamic theory for the velocity is multiplied by an extra $v^2$, compared to mass and Laplacian terms, so that when $v_0\to0$, the coupling $J$ must diverge as $1/v_0^2$ to keep finite the ferromagnetic interaction of the equilibrium theory.


\subsection{Hydrodynamic break down for large speed}

In our work we consider two values of the speed, $v_0=0.05$ (low activity) and $v_0=0.2$ (high activity). As we have seen, the activity coupling constant grows as the square of the speed, hence this change of a factor $4$ is sufficient to produce the crossover we wanted to test. However, $0.2$ may seem still quite a low value, hence one may ask what happens by increasing the speed even more and why we did not consider very large values of $v_0$ to test our theoretical results. The answer is that, although one is of course entitled to run the microscopic theory at any value of $v_0$, the hydrodynamic description of such theory breaks down when $v_0$ becomes too large. More precisely, in order for hydrodynamics to hold, one must have $v_0 \ll L$ (remember that the time step is set equal to $1$ in the model, which accounts for the weird dimensional nature of this relation), which, given our systems' sizes, limits the highest values of $v_0$ to the range $0.2-0.3$. The reason for this is the following.

In general \cite{landau_kinetic}, the continuum description of a microscopic theory is applicable only if the Knudsen number, $\mathrm{Kn}$, that is the ratio between mean free path, $\sigma$, and macroscopic length scale, $L$, is small, 
\beq
\mathrm{Kn} = \frac{\sigma}{L} \ll 1
\label{knu}
\eeq
In the RG context, the main idea is that particles must undergo many interactions within the coarse-graining volume, in order for coarse-graining to make sense at all; because such volume must be much smaller than the entire system, equation \eqref{knu} is a necessary condition for the field theory to make sense. In the case of the Vicsek model, due to the time-discrete nature of the dynamics with $\Delta t=1$, the mean free path for large speed can be approximated as, $\sigma \sim v_0 \Delta t = v_0$, so that the range of validity of a hydrodynamic description of the Vicsek model is limited by the condition, 
\beq
v_0 \ll L \ .
\eeq
 If we violate this conditions, the continuum description breaks down, starting with the very concepts of correlation length and relaxation time. At the intuitive level, it is clear that when $v_0 \sim L$ the very notion of local interaction disappears, and the complete rewiring of the interaction neighbourhood of each particle at each time step makes a standard, albeit off-equilibrium, statistical description of the problem quite problematic.


\subsection{Static correlation function and correlation length}
When the Vicsek model presents a continuous phenomenology, it is possible to compute the usual quantities in classical statistical physics to describe the phase transition \cite{cardy1996scaling}. However, due to the out-of-equilibrium nature of the self-propelled particles, all the correlation functions involve fluctuations of the order parameter velocity at time $t$ computed with respect to the velocity of the center of mass at the same time \cite{cavagna2018physics}, namely:
\begin{equation}
\delta \bm v_i (t) = \bm v _i - \frac 1N \sum_{i=1}^N  \bm v_i 
\label{deltav}
\end{equation}
This definition implies the sum rule
\begin{equation}
\sum_i \delta \bm v_i(t) =0.
\label{sumRule}
\end{equation}

  Therefore, when we compute the spatial correlation function in $k$-space as
\begin{equation}
C(k)= \frac 1N \left\langle \sum_{ij}  \frac{ \sin (k r_{ij})}{k r_{ij}} \delta \bm v_i (t) \cdot \delta \bm v_j(t) \right\rangle ,
\end{equation}
where the brackets $\langle  \ \rangle$ indicate an average in time, we have to take into account the fact that
the correlation function computed at $k=0$ is null and thus the usual definition for the susceptibility $\chi = \int d^d r \ C(r) = 0$ loses its meaning.
As a consequence, we identified as a measure of $\chi$ the maximum of the correlation $C(k)$ which occurs at a specific value of wave number $k_c$. 
This same value of wave number gave us an estimate of the correlation length $k_c = 1/ \xi$.  Indeed, when we compute the correlation functions we start considering very large $k$, or equivalently short distances and averaging over all the correlated pairs inside this region. It is clear then that the correlation function $C(k)$ increases when the wave number $k$ decreases since larger correlated regions are involved in the average and it reaches its maximum value when the length scale includes all the correlated pairs, namely at the correlation length $\xi = 1/k_c$. Decreasing more the wave number, the average starts to include not correlated regions and the function decreases till reaching the zero at $k=0$, as a consequence of the sum rule \ref{sumRule}. The inverse of $k_c$ is, therefore, a good estimate for the correlation length while the value of the correlation function computed at the same point can be interpreted as the susceptibility $\chi = C(k_c)$ \cite{cavagna2018physics}.

\subsection{Running simulations in the scaling region.}
For every size of the system with $N$ particles, we identified a finite-size critical point leaving the noise fixed at $\eta=0.45$ and varying the density of the system. For every value of the control parameter studied, we computed the susceptibility $\chi$ and we recognized as the critical point the value of density $\rho_c$ for which the susceptibility reached a maximum. At this value of density, we computed the correlation length $\xi$ as described above and we studied the dynamics at $k=1/\xi$. 
We considered systems of size $N=128, 256, 384, 512, 1024, 2048$ and we let  the length of the box vary in order to span the transition changing the mean nearest neighbour distance $r_1$. For each set of parameters we run from $10^5$ to $10^6$ steps (close to criticality) to equilibrate the system, then we collected $8$ independent samples of $5 \times 10^3$ steps to compute the required quantities. The metric interaction range is fixed to $r_c =1$.

\subsection{Estimate of the relaxation time}

The dynamic correlation function is defined as \cite{cavagna2018physics}:
\begin{equation}
C(k,t) = \frac{1}{N} \left \langle \sum_{ij} \frac{\sin (k r_{ij}(t,t_0))}{k r_{ij} (t,t_0)} \delta \bm v_i (t_0) \cdot \delta \bm v_j (t_0+t) \right\rangle_{t_0}
\end{equation}
where $r_{ij}=|\bm r_i(t_0)-\bm r_j(t)|$, $\langle \ \rangle _{t_0} = 1/(T_{max}-t) \sum_{t_0=1}^{T_{max}-t}$, and $T_{max}$ is the length of the simulation. Normalizing these functions at their value at $t=0$, we define $\hat C(k,t)= C(k,t)/C(k,t=0)$. Averaging over different independent samples we obtain the average correlation functions, hence the characteristic time scale of these is determined from the condition
\begin{equation}
\frac{1}{2\pi} = \int_0^{\infty} \frac{1}{\tau} \sin \left( \frac{t}{\tau} \right) \hat C (k,t).
\end{equation}
The value of $\tau$ is, therefore, the time scale that realizes that half of the total integrated area of the dynamic correlation function in the frequency domain comes from the interval $-\omega_c < \omega < \omega_c$ with $\omega_c = 1/\tau$. This definition comes from dynamic critical phenomena literature \cite{HH1969scaling} and it is particularly handy since it is able to compute the relevant time scale of systems in different scenarios: both when the relaxation is dissipative or when it includes propagating modes.

\begin{figure}
\centering
\includegraphics[width=0.5 \textwidth]{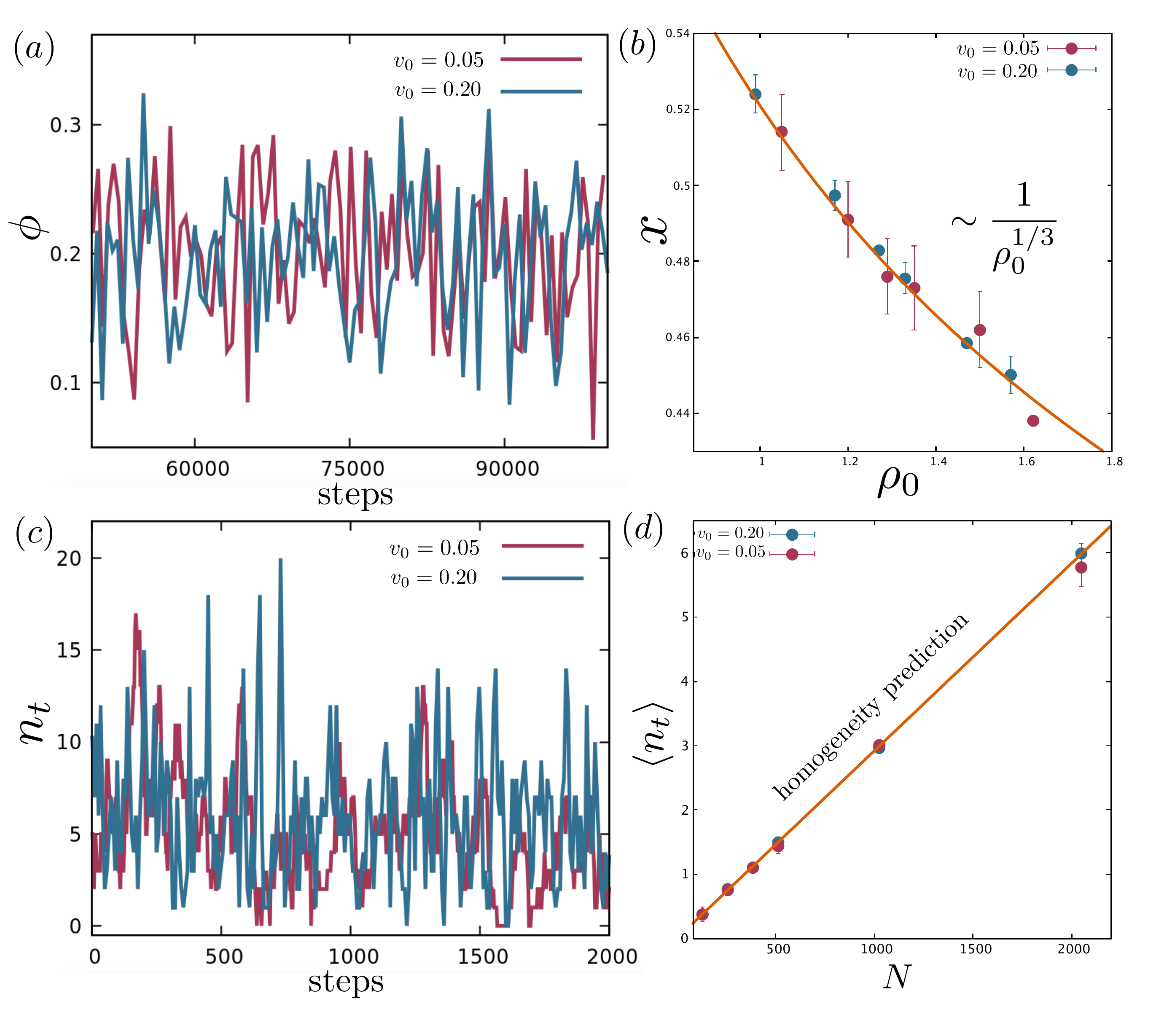}
\caption{{\bf Numerical simulations: test of homogeneity.} Analysis of homogeneity for data sets with $v_0=0.05$ (red) and $v_0=0.2$ (blue). (a) Temporal series of polarization $\phi$ for both activities and for size $N=2048$ at the critical point. (b) The mean first neighbour distance is computed for each size $N=128, 256, 384, 512, 1024, 2048$, it scales as $x \sim \rho_0^{1/3}$ as predicted for homogeneous systems. (c) Temporal series of $n_t$: the number of particles entering in a box of size $l_0=L/7$ and centered in the middle of the bigger box of size $L$ for $N=2048$. (d) Average on time of $n_t$ for all the sizes $N$ and speed: this quantity scales linearly with $N$, as predicted for homogeneous systems $\langle n_t \rangle = N/7^3$.
}
\label{homo}
\end{figure}

\subsection{Tests of homogeneity} 

For all the simulations collected, we looked at the temporal series of the polarization $\phi = \frac 1N | \sum \bm v_i |$, monitoring  jumps in its average value as a trace of higher density structures in the ensemble (Fig. \ref{homo}a). Moreover, we did additional basic tests  to check the spatial homogeneity condition: first, we analyzed the mean first neighbour distance $x = r_1/r_c$ that turns out to properly scale as $x \sim 1/ \rho^{1/3}$, as it should be in a homogeneous system (Fig. \ref{homo}b); second, we localized a box of size $l_0 = L/7$ inside the bigger box of the system and we counted how many particles  $n_t$ entered in the small box during the simulations. We looked at its temporal series and then we averaged it in time obtaining $\langle n_t \rangle$ \cite{ginelli2016physics}. We expected, for systems without travelling bands or aggregates of particles, $\langle n_t \rangle$ to fluctuate around a mean value and to scale in size as:
\begin{equation}
\langle n_t \rangle \sim \rho_0 \  l_0^3 
\label{homoeq}
\end{equation}
where $\rho_0$ is the average density \cite{ginelli2016physics}. In Fig.~\ref{homo} we report data for the two values of activity and the sizes $N$ analyzed: in Fig.~\ref{homo}c we show two temporal series of $n_t$ for $N=2048$ and  in Fig.~\ref{homo}d we see that  $\langle n_t \rangle$  follows very well the prediction of  \eqref{homoeq}. Indeed, using information on $l_0$ and of $\rho_0 = N/ L^3$, we obtain a linear trend in size as $ \langle n_t \rangle \sim  N/7^3$.

\subsection{Statistical tests about the fit of the critical exponent}

To carefully discriminate between the dynamic critical exponents of the two different fixed points, we performed three basic statistical tests on our simulations.

\begin{figure}
\centering
\includegraphics[width=0.5 \textwidth]{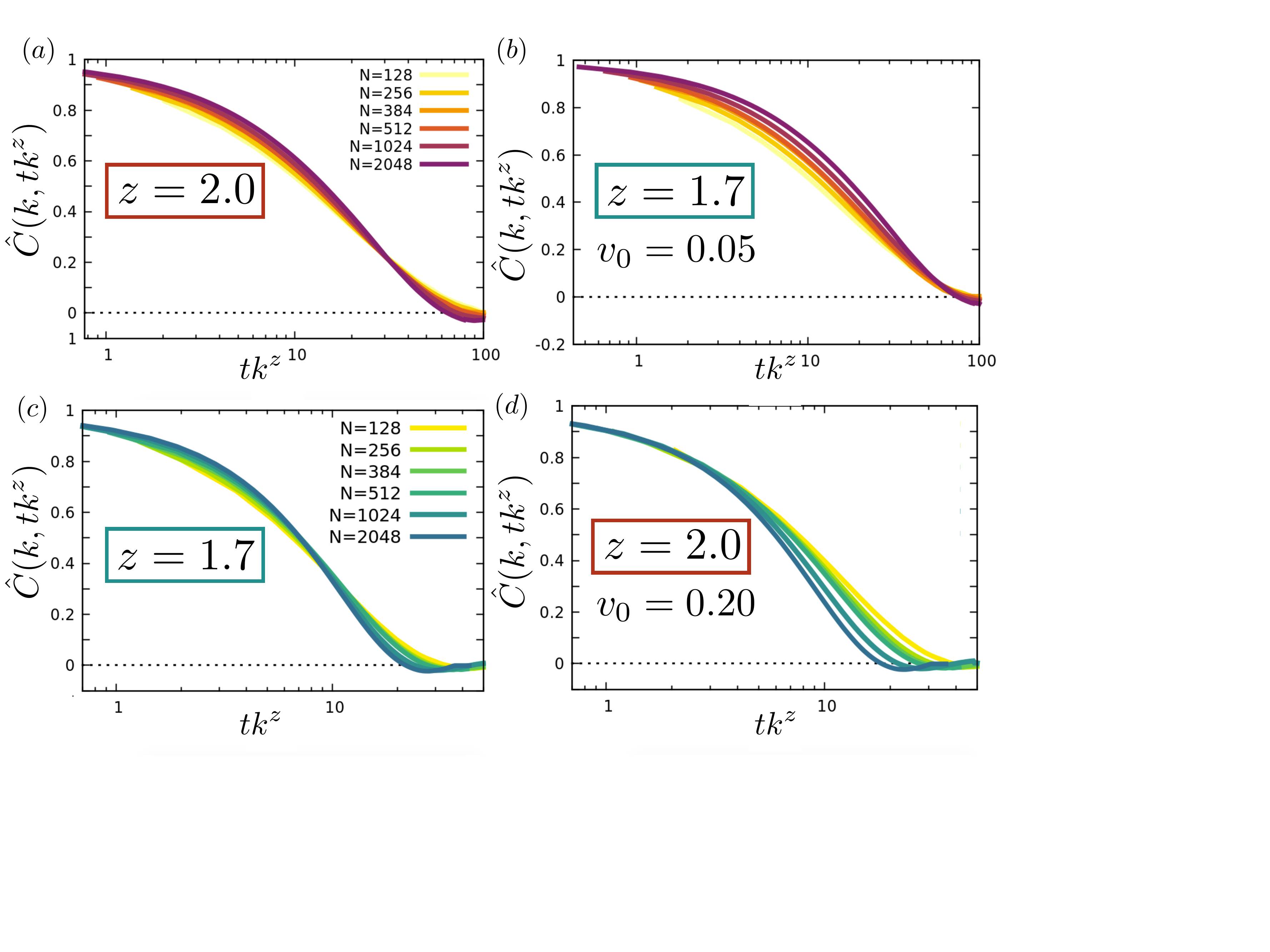}
\caption{{\bf Numerical simulations.} $(a)$, $(c)$ Scaling of dynamic correlation functions $\hat C(k,t)$ using values of $z$ extrapolated by the linear fits of Fig.\ref{simu}:  $(a)$, $v_0=0.05$ with $z =2.0$  and $(b)$, $v_0=0.20$ with $z = 1.7$. Panels $(b)$, $(d)$ same correlation functions reversing the scaling procedure:  $(c)$, $v_0=0.05$ with $z =1.7$ and $(d)$, $v_0=0.05$ with $z = 2.0$. The scaling hypothesis is not well verified for $(b)$ and $(d)$.
}
\label{simu1}
\end{figure}

Test 1: We rescaled the dynamic correlation functions for low speed with $z=1.7$ and the ones for high speed with $z=2$, reversing the scaling procedure showed in the main text. The result is reported in Fig.\ref{simu1} where we compare the rescaling of the $\hat C(k,t)$ using the value of $z$ obtained by the linear fit in the plane $(\ln \xi, \ln \tau)$, with the scaling realized using the exponent of the other fixed point. A discrepancy is visible and this confirms that dynamic correlation functions for speed $v_0=0.05$ scale better with the exponent $z=2$ while the functions for higher speed $v_0=0.2$ scale properly for $z=1.7$.

Test 2: We examined the consistency of the values of $z$ extrapolated by linear fits of the data with the theoretical expectations.
First of all, for each value of speed $v_0$ we fit the data on the plane $(\ln \xi, \ln \tau)$ with the simple linear function $f(x)=mx+c$. Since we know that $\tau \sim \xi^z$, the slope $m$ is the value of $z$ we are looking for. We obtained:
\begin{itemize}
\item $v_0=0.05$: $z_{\mathrm{sim}} = 2.10 \pm 0.04$
\item $v_0=0.20$: $z_{\mathrm{sim}} = 1.65 \pm 0.03$
\\
\end{itemize}

The value of the exponent $z_{\mathrm{sim}}$ is affected by an uncertainty $\sigma$, which is computed assuming that data on correlation time belong to a gaussian distribution and that their standard deviation can be determined with the least squares method. Then, we performed a basic hypothesis test to verify the compatibility with the expected theoretical values $z_{\mathrm{th}}$. We computed the variable
\begin{equation}
t = \frac{ z_{\mathrm{sim}} - z_{\mathrm{th}}}{\sigma}
\end{equation}
which measures the distance between the theoretical and the simulation value of $z$ in units of uncertainty. If $|t| > 3$, the probability that the value extracted from simulations is compatible with the theoretical one is less than $1 \%$. We tested:
\begin{itemize}
\item $v_0=0.05$: $z_{\mathrm{th}} = 2.0 \to t=2.5$, \ \ \ \it{consistent};

\qquad \ \ \qquad  $z_{\mathrm{th}} = 1.7 \to t=10.0$, \ \ \it{not consistent}.
\item $v_0=0.20$: $z_{\mathrm{th}} = 2.0 \to t =-11.7$,  \it{not consistent};

\qquad \ \ \qquad $z_{\mathrm{th}} = 1.7 \to t= - 1.7 $, \it{ consistent}.
\end{itemize}

This result  validates the thesis we found the dynamic crossover in the Vicsek model. 

Test 3: We evaluated the goodness of the linear fits of Fig.\ref{simu} using theoretical values of the dynamic critical exponents as fixed slope. To achieve this, we carried out a $\chi^2$-test. For each data set with activity $v_0$, we performed a linear regression using both values of $z_{\mathrm{th}}=2.0,1.7$ and extrapolating only the intercept $c$ of the linear function from data.  With  uncertainties on $\ln \tau$ derived by the least square method, as in the point above, we computed the standard $\chi^2$ for $N-1$ degrees of freedom for both the analysis. If this variable results larger than $11$, we can say that the probability that the fit is compatible with data is less than $5\%$, hence

\begin{itemize}
\item $v_0=0.05$: $z_{\mathrm{th}} = 2.0 \to \chi^2=9.98$, \it{consistent};

\qquad \ \ \qquad  $z_{\mathrm{th}} = 1.7 \to \chi^2=415$, \ \it{not consistent}.

\item $v_0=0.20$: $z_{\mathrm{th}} = 2.0 \to \chi^2 =131$,  \ \it{not consistent};

\qquad \ \ \qquad  $z_{\mathrm{th}} = 1.7 \to \chi^2= 7.04 $, \it{ consistent}.
\end{itemize}
Once again, this analysis confirms the result of this work, namely that, increasing the activity of the self-propelled particles in the standard Vicsek model, we can numerically verify the dynamic crossover from the equilibrium to the out-of-equilibrium dynamic universality class.

\end{document}